\documentclass[english,nofootinbib,notitlepage]{mn2e}

\usepackage[T1]{fontenc}
\usepackage[latin1]{inputenc}
\usepackage{natbib}

\usepackage{amsmath}
\usepackage{amssymb}
\usepackage{babel}
\usepackage{array}
\usepackage{placeins}

\usepackage{ifpdf}
\ifpdf   
    \usepackage[pdftex]{graphicx}
    \usepackage[hyperfootnotes=false,pdfstartview=,colorlinks=false,linkcolor=BrickRed,citecolor=blue,urlcolor=blue,bookmarks=false,pagebackref]{hyperref}    \DeclareGraphicsExtensions{.png,.jpg,.pdf}

\title[Simultaneous constraints on bias, normalization and growth index]{Simultaneous constraints on bias, normalization and growth index
through power spectrum measurements}

\author[Di Porto et al.]{Cinzia Di Porto$^{1,2}$, Luca Amendola$^{2}$, Enzo Branchini$^{3}$\\
$^{1}$INAF-Osservatorio Astronomico di Bologna, Via Ranzani 1, 40127, Bologna, Italy and INFN Sezione di Bologna\\
$^{2}$Institut f$\ddot{\textrm{u}}$r Theoretische Physik, Universit$\ddot{\textrm{a}}$t Heidelberg, Philosophenweg 16, 69120 Heidelberg, Germany\\
$^{3}$Dipartimento di Fisica ``E. Amaldi'', Universit\'a degli Studi ``Roma Tre'', via della Vasca Navale 84, 00146, Roma, Italy,  
\\INFN Sezione di Roma Tre and  INAF, Osservatorio Astronomico di Brera, Milano, Italy}

\begin{document}

\date{Submitted:}

\pagerange{\pageref{firstpage}--\pageref{lastpage}} \pubyear{2012}
\maketitle

\label{firstpage}

\begin{abstract}
In this Letter we point out that redshift surveys can break
the degeneracy between the galaxy bias, the power spectrum normalization,
$\sigma_{8,0}$ and the growth factor, without the need for external
information by using a simple and rather general 
 parametrization for the growth rate,
 the well known $\gamma$ parametrization and measuring the power spectrum at least at two different redshifts. We find that in 
next-generation surveys like Euclid,
$\sigma_{8,0}$ and $\gamma$ can be measured to within $1\%$ and $5\%$,
respectively, while the bias $b(z)$ can be measured to within $1-2\%$
 in each of 14 equal-width redshift bins spanning $0.7\leq z
\leq 2$.
\end{abstract}
\maketitle

\section{Introduction}

The issue of constraining the cosmological parameters by employing
the 3D galaxy power spectrum as a summary statistic of the matter
density perturbations has been widely explored
in the last few years. Indeed, the power spectrum estimated from 
a redshift survey can give a wealth of information through
its shape, amplitude and radial anisotropy induced by peculiar velocity-driven
redshift space distortions (RSD). In principle the shape can be used
to constrain background quantities such as the dark energy equation
of state and the geometry of the system that are related to the expansion
history of the universe. The amplitude and RSD of galaxy clustering
can constrain the growth of cosmological perturbations and the biasing
relation between galaxies and mass, i.e. the relation between the spatial
distribution of galaxy and the underlying mass density field.

Constraining the growth is essential in order to discriminate dark
energy models with the same background properties but different physical
origins. This makes the power spectrum a powerful tool in discriminating
standard dark energy models from modified gravity theories.
However, sharp discrimination is possible only if galaxy bias can
be estimated precisely or, at the very least, marginalized over efficiently.

Even though the work of \cite{kaiser87} made it clear that one of
the most promising ways to determine the fluctuation growth is to exploit
the redshift distortion effect (see \cite{hamilton98} for a review),
in the first pioneering works \citep{seo03,seo07} dealing with Fisher
matrix formalism as a tool to constrain cosmological parameters with
redshift surveys, the redshift distortions were in fact considered
a sort of ``noise'' disturbing the measure of the baryon acoustic
oscillations peaks. The attitude was  to marginalize over the
RSD and  the growth. Later on, it was realized that  information
contained in the redshift distortions could
tighten constraints over cosmological parameters \citep{amendola05,wang08,linder07}
and RSD began to be considered as a standard, additional probe \citep{guzzo08},
able to constrain both the growth rate and the dark energy 
responsible for  the accelerated expansion of  the universe. The growth factor itself
began to be regarded as useful information to be exploited rather than to be
marginalized over \citep{amendola05,wang10}.

However, as noticed in several works \citep{wang08,percival09,song09},
the degeneracies between the growth factor $G(z)$, the (linear) bias
$b(z)$ and the power spectrum normalization $\sigma_{8,0}\equiv\sigma_{8}(z=0)$
do not allow us to constrain all three quantities simultaneously. One can
only measure combinations such as $f(z)\sigma_{8}(z)=f(z)G(z)\sigma_{8,0}$,
$b(z)\sigma_{8}(z)=b(z)G(z)\sigma_{8,0}$ or  $\beta(z)\equiv f(z)/b(z)$
(e.g. \cite{ross06,guzzo08,blake11}), 
where $f(z)$ is the growth rate, related to the growth factor via $f(z)=d\ln G(z)/d\ln a$
and $a=(1+z)^{-1}$ is the scale factor. 
The growth rate was originally parametrized by \cite{peebles80} as $f(z)= 
\Omega_m(z)^{0.6}$ in a matter-dominated cosmology. This expression was
later generalized as $f(z)=\Omega_m(z)^{\gamma}$ to accurately describe the growth rate in a 
variety of cosmological models ranging from Dark Energy models to non-standard gravity theories \citep{lahav91,wang98,amendola04,linder05,polarski07}.
This parametrization obviously implies $G(z)=\int\Omega_{m}^{\gamma}/(1+z)\, dz$.


Several authors have adopted different strategies to break, bypass or ignore
the degeneracy between $G(z)$, $b(z)$ and $\sigma_{8,0}$ depending on the parameters they were interested in constraining. 

\begin{itemize}
\item  In general one fixes  (or assumes external priors for)  one of the parameters and then estimate the others. 
Different authors have fixed the clustering amplitude $\sigma_{8,0}$ to the value estimated
from the cosmic microwave background (CMB) data to constrain $f(z)$ and $b(z)$ (e.g. \citealt{diporto11}) or the
growth parameter $\gamma$ (e.g. \citealt{diporto11,belloso11}) or both
the bias and the growth (e.g. \citealt{diporto11}). Instead of fixing $\sigma_{8,0}$ to a single value, some authors have assumed CMB priors for this parameter to constrain $f(z)$ (e.g. \citealt{guzzo08}) or $\gamma$ and $b(z)$ \citep{gaztanaga11}. 
\item One can assume that General relativity holds and fix $\gamma$ to the $\Lambda$ cold dark matter ($\Lambda$CDM)
value (e.g. \citealt{seo03,amendola05,ross06,wang10}). For example \cite{hawkins03} and \cite{cole05} fix $\gamma$ to estimate the bias while \cite{percival04}
assume a further prior on $\Omega_m$ to determine
$\sigma_{8,0}$.
\item Alternatively, one can consider some measurable combination of the above parameters that
can also discriminate models efficiently. A popular choice is the
{``}mass weighted{''} growth rate $f(z)\sigma_{8}(z)$ which
provides a good test for dark energy models (e.g. \citealt{song09,white09,percival09,blake11,carbone10,wang10})

\end{itemize}
It is a fact that galaxy bias, growth and clustering amplitude cannot
be independently estimated from the observed $P(k)$ alone \citep{percival09}
and yet effective constraints can be placed provided that these quantities
could be accurately parametrized under fairly general hypotheses.
Although the idea is somewhat implicit in some of the works quoted
above, it seems to us it has not been clearly
pointed out nor discussed thoroughly anywhere.
The scope of this work is to make it explicit with a simple proof,
that we present in Sec. \ref{sec:method}. In Sec. \ref{sec:fm} we
present the Fisher matrix method that we employ in order to obtain
the aforementioned constraints on parameters. Results are presented
in Sec. \ref{sec:results} and finally in Sec. \ref{sec:conclusions}
we draw our conclusions.

\section{Lifting Parameter Degeneracy} \label{sec:method}

Let us consider a measurement of the galaxy power spectrum in redshift
space at different epochs, i.e. in different redshift bins $z$. In
the linear regime, the shape of the power spectrum and its redshift
distortions can be modeled as follows \citep{seo03,seo07}

\begin{eqnarray}
P_{\textrm{obs}}(z;k,\mu) &\!\!\!\!\! =&\!\!\!\!\! \frac{D_{F}^{2}(z)H(z)}{D^{2}(z)H_{F}(z)}G^{2}(z)b^{2}(z)\sigma_{8,0}^{2}\left[1+\frac{f(z)}{b(z)}\mu^{2}\right]^{2}P_0\nonumber\\
 &&\!\! +P_{\textrm{s}}(z)\nonumber\\
&\!\! \!\!\!\equiv &\!\! \!\!\!C(z)G^{2}(z)B^{2}(z)\left[1+R(z)\mu^{2}\right]^{2}P_{0}+P_{\textrm{s}}(z)\label{eq:spectrum}
\end{eqnarray}
 where $B(z)\equiv b(z)\sigma_{8,0}$, $R(z)=\frac{f(z)\sigma_{8,0}}{B(z)}$,
$\mu$ is the direction cosine and $P_{0}\equiv P(k,z=0)$ is the
linear power spectrum at the present epoch. The factor $C(z)\equiv {D_{F}^{2}(z)H(z)}/({D^{2}(z)H_{F}(z))}$, where $D(z)$ is the angular diameter distance and $H(z)$ is the Hubble parameter, takes into account the difference in comoving
volume between the fiducial cosmology - the one we use to convert observed
redshifts into distances - (subscript $F$) and the true cosmology. Finally, we model the shot noise contribution
as an additional factor $P_{s}(z)$.

However, let us notice that, since we are not interested in constraining $D(z)$ and $H(z)$ in every redshift bin, but only the parameters they depend on, these functions and their 
combination $C(z)$ are not considered further in this Letter.


Let us assume that the growth rate can be modelled as $\Omega_{m}(z)^{\gamma}$.
The growth index $\gamma$ needs not to be constant. The simple parametrization 
$\gamma(z)=\gamma_{0}+\gamma_{1}z/(1+z)$ has been shown
to reproduce the time dependence of $\gamma$ in a variety of non
standard gravity models \citep{fu09}.
For the sake of simplicity we consider in this Section the case of constant $\gamma$, but our conclusion remains valid
also with  time-dependent growth index, as we argue below.

Cosmological parameters are determined by fitting Eq.~(\ref{eq:spectrum})
to the observed galaxy power spectrum. From the spectral shape in the linear
regime, $P_{0}$, one can determine those parameters that describe the
expansion history $H(z)$ (and then $D(z)$) ($h$, $\Omega_{m,0}$, $\Omega_{DE}$,
$w_{0}$, $w_{1}$, etc...). From the amplitude of the power spectrum
one determines the combination $A(z)\equiv G(z)B(z)$. 
If this analysis can be performed at two (or more) redshifts and 
the growth rate is modelled as $f=\Omega_{m}(z)^{\gamma}$, then 
the degeneracy can be broken. Let us prove it for the simple case in which 
the power spectrum has been measured at two redshifts
$z_{1}$ and  $z_{2}$.
In this case, one determines $A$ and $R$ at two different epochs,
$A_{1},\, R_{1},\, A_{2},\, R_{2}$ and can solve the linear system
\begin{eqnarray}
A_{1} & = & G_{1}(\gamma)B_{1}\\
A_{2} & = & G_{2}(\gamma)B_{2}\\
R_{1} & = & \frac{\Omega_{m,1}^{\gamma}\sigma_{8,0}}{B_{1}}=\frac{\Omega_{m,1}^{\gamma}\sigma_{8,0}G_{1}(\gamma)}{A_{1}}\\
R_{2} & = & \frac{\Omega_{m,2}^{\gamma}\sigma_{8,0}}{B_{2}}=\frac{\Omega_{m,2}^{\gamma}\sigma_{8,0}G_{2}(\gamma)}{A_{2}}\;.
\end{eqnarray}
 By assumption $\Omega_{m,z_i}$ has been already determined because
it depends only on the background parameters. Then the ratio of the
last two equations yields 
\begin{equation}
\frac{R_{1}A_{1}}{R_{2}A_{2}}=\left(\frac{\Omega_{m,1}}{\Omega_{m,2}}\right)^{\gamma}\frac{G_{1}(\gamma)}{G_{2}(\gamma)}
\end{equation}
 where the only unknown is $\gamma$ and therefore we can solve for
it. Then, from the above equations, one estimates $\sigma_{8,0}$
and $b(z)$ as 
\begin{eqnarray}
\sigma_{8,0} & = & \frac{A_{i}R_{i}}{G_{i}(\gamma)\Omega_{m,i}^{\gamma}}\ \ \ \ {\rm and}\ \ \ \ \ b(z_{i})=\frac{\Omega_{m,i}^{\gamma}}{R_{i}}
\end{eqnarray}

 We notice that if the power spectrum is measured in three redshift bins, then 
 one can also constrain the time dependence  of the growth index 
 $\gamma(z)=\gamma_{0}+\gamma_{1}z/(1+z)$.
This procedure assumes that the size of the bins
and the number densities of galaxies within are large enough
to provide effective constraints to avoid that
parameters' degeneracy, although broken
in principle, can reappear due to poor statistics. As we show in the next section,
this is, however, not of concern
for large surveys as Euclid.
The proof that the degeneracy can be broken 
can be also obtained using the Fisher matrix formalism
\citep{fisher35,tegmark97r}.
If one assumes $ f=\Omega_{m}(z)^{\gamma}$,
a constant growth index  and the model power spectrum 
of Eq.~(\ref{eq:spectrum}), then the derivatives of the spectrum with
respect to $b(z),\,\gamma$ and $\sigma_{8,0}$ 
\begin{eqnarray}
\frac{\partial\ln P_{\textrm{obs}}}{\partial\ln b_{z_{i}}} & = & 2-2\frac{\Omega_{m}(z_{i})^{\gamma}\mu^{2}}{b_{z_{i}}+\Omega_{m}(z_{i})^{\gamma}\mu^{2}}\\
\frac{\partial\ln P_{\textrm{obs}}}{\partial\gamma} & = & 2\frac{\partial\ln G}{\partial\gamma}+2\frac{\Omega_{m}(z_{i})^{\gamma}\ln\Omega_{m}(z_{i})\mu^{2}}{b_{z_{i}}+\Omega_{m}(z_{i})^{\gamma}\mu^{2}}\\
\frac{\partial\ln P_{\textrm{obs}}}{\partial\sigma_{8,0}} & = & \frac{2}{\sigma_{8,0}}\,
\end{eqnarray}
 are not degenerate because of their different dependence from $z$
and $\mu$. Dropping the $\gamma$ parametrization, i.e. treating
$G(z)$ as an extra free parameter for every bin, would not allow us to
lift the parameter degeneracy and a singular Fisher matrix would result.
We note that here we assumed the bias to be scale-independent which
should be true on the large, linear scales we are focusing on here. A
possible scale dependence would modify the shape of the spectrum and
should be accounted for to avoid systematic errors in the estimate
of the background cosmology parameters.
Furthermore, a scale-dependent bias could make the Fisher matrix degenerate again and cause the parameter degeneracies to return.
Clearly, in order to break the degeneracy in presence of a bias scale
dependence, one should
parametrize it as, e.g., a simple power law of $k$, employing a number
of parameters smaller
than the number of distinct redshift bins. This issue will be studied in a future work.

\section{Fisher matrix forecasts}

\label{sec:fm}

We now apply the strategy outlined above to a specific measurement
and adopt the Fisher matrix approach to evaluate the expected errors
on the measured parameters. For this purpose we choose a reference
cosmological model and run CAMB~\citep{lewis00} to obtain 
the linear power spectrum in real space $P_{0}$. 

Then, the model power  spectrum in Eq.~\ref{eq:spectrum} depends on a number of cosmological parameters, listed in Tab.~\ref{tab:parameters}.
In the upper part we list parameters that do not depend on redshift.
Redshift-dependent parameters are listed in the bottom part of the
table. In our application we consider the more general case 
of a time-dependent growth index  $\gamma(z)=\gamma_{0}+\gamma_{1}z/(1+z)$.
We adopt a slightly unconventional fiducial $\Lambda$CDM
model with $\gamma_{0}=0.545,\ \ {\rm and}\ \ \gamma_{1}=0$ (to
be compared with the concordance $\Lambda$CDM values $\gamma_{0}=0.556$,
$\gamma_{1}=-0.018$~\citep{fu09}) and  allow for a
time-dependent equation of state for the Dark Energy with $w_{0}=-0.95$
and $w_{1}=0$.

In this work we do not consider a possible smearing of the 
wiggles associated with baryonic
acoustic oscillation (BAO) \citep{seo07}. Further, we ignore non-linear
effects that modify the spectral shape and the RSD pattern
on small scales (e.g.
\cite{blake11}) by limiting our analysis to small wavenumbers $k\leq k_{\textrm{max}}\equiv \textrm{min}[k_{\textrm{cut}}, k_{\textrm{lin}}(z)]$, where $k_{\textrm{cut}}$ is set to $0.2\,h\,\textrm{Mpc}^{-1}$.
The value for 
$k_{\textrm{lin}}(z)$ is set by requiring that the variance
in cells $\sigma^{2}(k_{\textrm{lin}},z)=0.25$ (for example, at $z=0.7$
we have $k_{\textrm{lin}}\sim0.16\, h\,\textrm{Mpc}^{-1}$).

Having assumed our model power spectrum and  
set the parameter values of the fiducial model, we compute the elements of the Fisher matrix as in \cite{tegmark97}
by integrating over all modes below $k_{\textrm{max}}$. The result
depends on the characteristics of the galaxy redshift catalogue used to compute
the power spectrum. More precisely, it depends on the volume of the
survey and on the galaxy redshift distribution $dN/dz$. In this
work we take, as a reference case, the Euclid redshift survey as specified
in the {\it Red Book} \citep{euclid_rb} and at the website
http://www.euclid-ec.org/.
This survey will span a broad redshift
range $0.7\leq z \leq 2$ that we split in 14 bins of $\Delta z=0.1$.
The expected sky coverage is
 $15000$ deg$^{2}$ . The expected $ dN/dz$
is given in the aforementioned {\it Red Book}, while the fiducial values for the bias in each redshift bin, for Euclid galaxies, are derived from \cite{orsi10}.

\begin{table}
\begin{minipage}{84mm}%
\caption{\label{tab:parameters} List of the parameters that characterize the
background cosmology and the model power spectrum. They are used in
the Fisher matrix analysis. Redshift dependent parameters are listed
in the lower part of the table.}

\begin{tabular}{|>{\raggedright}m{4.2cm}>{\centering}m{1cm}>{\centering}m{2.0cm}|}
\hline 
Redshift-independent parameters  &  & fiducial values\tabularnewline
Reduced total matter density  & $\Omega_{m,0}h^{2}$  & $0.271\cdot(0.703)^{2}$\tabularnewline
Reduced baryon density  & $\Omega_{b,0}h^{2}$  & $0.045\cdot(0.703)^{2}$\tabularnewline
Curvature density  & $\Omega_{k}$  & 0\tabularnewline
Hubble constant at present  & $h$  & 0.703\tabularnewline
Primordial fluctuation slope  & $n_{s}$  & 0.966\tabularnewline
Dark energy eq. of state  & $w_{0},w_{1}$  & -0.95, 0\tabularnewline
Power spectrum normalization  & $\sigma_{8,0}$  & 0.809\tabularnewline
\textit{$\gamma$-parameterization} parameters  & $\gamma_{0}$, $\gamma_{1}$  & 0.545, 0\tabularnewline
 &  & \tabularnewline
Redshift-dependent parameters  &  & \tabularnewline
(in 14 $z$-bins) &  & \tabularnewline
Shot noise  & $P_{s}$  & 0\tabularnewline
Bias  & $\log b$  & Derived from \cite{orsi10} \tabularnewline
\hline 
\end{tabular}%
\end{minipage}
\end{table}

\section{Results}

\label{sec:results}

\begin{figure*}
\centering
\begin{minipage}{160mm}

\begin{centering}
 \includegraphics[height=7.5cm]{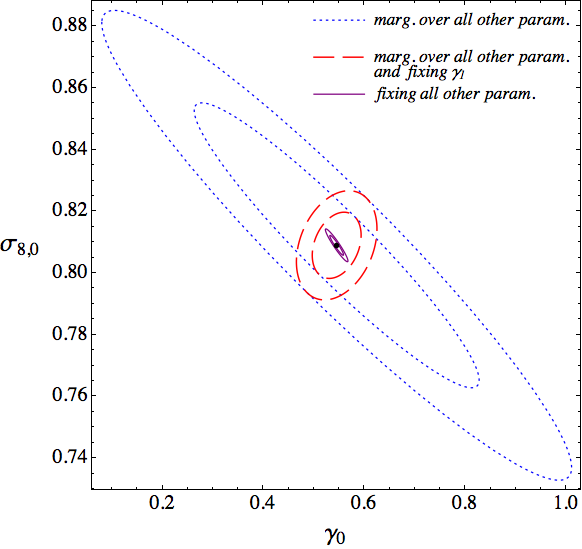}$\quad$ \includegraphics[height=7.5cm]{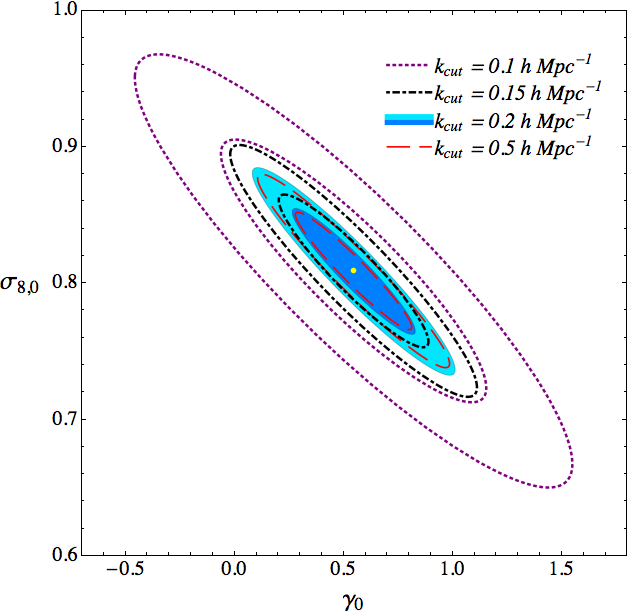}
 
 \par\end{centering}

\caption{\label{fig:gamma_s8_cplot} Left-hand panel: contours
at 68\% and 95\% of probability for the parameters $\gamma_{0}$ and
$\sigma_{8,0}$ when $i)$ we marginalize over all other parameters
(blue, dotted ellipses); $ii)$ we marginalize over all other parameters
after fixing $\gamma_{1}$ (red, long-dashed ellipses) and $iii)$
we fix all other parameters. The last case is represented by the very
small purple, solid ellipses nearly overlapping on the black dot which
marks the fiducial model. The marginalized 1$\sigma$ errors for $\sigma_{8,0}$ are 0.03, 0.007, 0.002 for the 3 cases, respectively while those for $\gamma$ are 0.19, 0.03, 0.009. Right-hand panel: contours at 68\% and 95\% of probability for the parameters $\gamma_{0}$ and $\sigma_{8,0}$ varying $k_{cut}$.}
\end{minipage}
\end{figure*}

Our analysis allows us to set simultaneous constraints
on $b(z),\sigma_{8}$ and $\gamma$, when the power spectrum is measured at two or more redshifts. Tabs. \ref{tab:Errors-on-cosmological},\ref{bias_errors}
display the expected $1\sigma$ uncertainties on these (and other) parameters, listed in the first
row. 
Errors on each parameter are obtained after marginalizing over all other parameters in the table. Therefore 
they do not coincide with the length of the ellipses' axis shown 
in Fig.~\ref{fig:gamma_s8_cplot} which represent joint, rather than marginalized, probabilities.
Table~\ref{tab:Errors-on-cosmological}
shows the uncertainties on the redshift independent parameters listed
in the top row. 
In the second row the values refer to 1$\sigma$ uncertainties
computed marginalizing over all other parameters while
in the bottom row  they are obtained through marginalization after fixing $\gamma_{1}$
in order to compare our results to other, similar analysis. In Table~\ref{bias_errors}
we list the expected uncertainties on the bias parameter estimated at the
different redshifts specified in the top row. The meaning of 
the second and third rows is the same as in Table~\ref{tab:Errors-on-cosmological}.

In the case of a time-dependent growth index, uncertainties
on the parameters are remarkably small
$\Delta \sigma_{8}=0.03 \;, \Delta \gamma_0 =0.19 \; {\rm and} \; b(z)=2-4\%$
depending on the redshift. They decrease
even further if $\gamma$ is assumed to be constant 
($\Delta\sigma_{8}=0.007$, $\Delta\gamma=0.03$ and $\Delta b(z)=1-1.7\%$).
In Fig. (\ref{fig:gamma_s8_cplot}) we plot the confidence regions
of $\gamma_{0}$ and $\sigma_{8}$. Contours refer to 68\% and 95\% joint probability
levels. The blue, dotted ellipses refer to the case in which
we marginalize over all parameters, including $b(z)$. The contour levels
shrink considerably when one fixes the value of $\gamma_{1}$ (red,
dashed) and become tiny in the case in which 
all cosmological parameters and the bias are fixed to their fiducial
values (we leave only the shot noise free to vary). This last case represents the ideal situation in which the available
complementary probes (e.g. SN Ia, CMB fluctuations, lensing etc) allow us 
to estimate all other parameters with very high precision. 

\begin{figure*}
\begin{minipage}{170mm}
 \includegraphics[height=7.5cm]{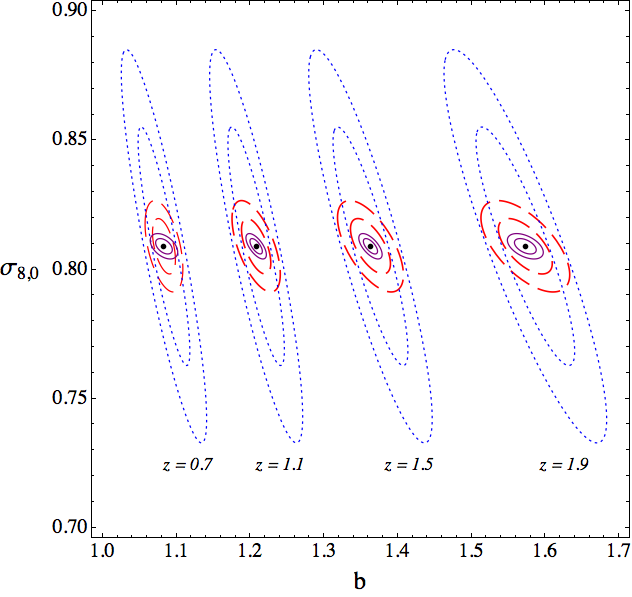}$\quad$ \includegraphics[height=7.5cm]{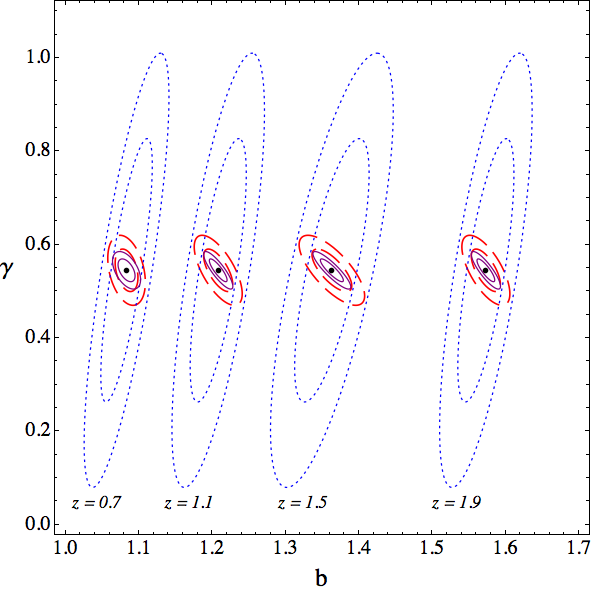}

\caption{\label{fig:gamma_s8_vs_b_cplot}Left-hand panel: the three sets of ellipses
show the contours at 68\% and 95\% of probability for the parameters
and $\sigma_{8,0}$ and $b(z)$ in 4 redshift bins when $i)$ we marginalize
over all other parameters (blue, dotted ellipses); $ii)$ we marginalize
over all other parameters after fixing $\gamma_{1}$ and $\Omega_{K}$
(red, long-dashed ellipse) and $iii)$ we fix all other parameters
(purple, solid ellipses). The black dots mark the fiducial values
for the parameters and are centered at the central values of each
redshift bin. Right-hand panel: as in left-hand panel, but for the parameters
$\gamma_{0}$ and $b(z)$.}
\end{minipage}
\end{figure*}

In Fig.~\ref{fig:gamma_s8_vs_b_cplot} we show the analogous confidence
contour levels in the $\sigma_{8}-b(z)$ (left-hand panel) and $\gamma-b(z)$
(right-hand panel) planes. As in the left-hand panel of Fig. \ref{fig:gamma_s8_cplot} ellipses
of different sizes refer to
marginalization over different sets
of parameters. The four sets of ellipses refer to
different redshift slices, as specified by the labels. We do not show all 14 redshifts
to avoid overcrowding. Keeping $\gamma$ constant and assuming a flat
universe allows to reduce errors on $\sigma_{8}$ significantly 
but has little effect on  galaxy bias uncertainties.

Finally, we have tested the sensitivity of our result to $k_{\textrm{cut}}$,
i.e. to the cut we have imposed to exclude nonlinear effects. 
Increasing $k_{\textrm{cut}}$ increases the number of $k$-modes used
in the analysis and reduces  statistical errors. However, if 
 $k_{\textrm{cut}}$  is too large, non-linear effects kick in and induce systematic errors. 
Decreasing $k_{\textrm{cut}}$ allows to apply linear theory 
but increases random errors. 
To find the best tradeoff one needs to estimate the power spectrum in some 
simulated galaxy catalog obtained from fully nonlinear N-body simulations.
However, we can obtain an order of magnitude estimate
 for the relative importance of non-linear effects by re-computing errors 
at  different values of $k_{\textrm{cut}}$ and check how much they change.
 
The right-hand panel of Fig. ~\ref{fig:gamma_s8_cplot} shows the result
of such exercise in the $\gamma_{0}-\sigma_{8}$ plane.
Ellipses filled with different shades of blue 
represent the reference case of $k_{\textrm{cut}}=0.2\, h\,\textrm{Mpc}^{-1}$ (and correspond to the blue, shorth-dashed ellipses in the left panel).
Pushing $k_{\textrm{cut}}$ up to $0.5\, h\,\textrm{Mpc}^{-1}$
(red, long-dashed) has a little effect, meaning that the decrease
in statistical errors does not justify the risk of 
introducing systematic errors driven by nonlinear effects.
Reducing $k_{\textrm{cut}}$ to $0.1\, h\,\textrm{Mpc}^{-1}$ has a quite dramatic
effect (purple, dotted ellipses). The large increase in the errors
reflects the fact that with this $k_{\textrm{cut}}$ one cuts off all but one
of the BAO wiggles. Finally, a value of $k_{\textrm{cut}}=0.15\, h\,\textrm{Mpc}^{-1}$
(black, dot-dashed ellipses) seems to represent a safer and yet acceptable
option.  However, the optimal choice of $k_{\textrm{cut}}$
depends on the underlying cosmological model and on the accurate modeling of nonlinear effects.

\begin{table*}
\centering %
\begin{minipage}[c]{160mm}%

\begin{tabular}{|>{\centering}p{0.8cm}|>{\centering}p{1cm}>{\centering}p{1cm}>{\centering}p{1cm}>{\centering}p{1cm}>{\centering}p{1cm}>{\centering}p{1cm}>{\centering}p{1cm}>{\centering}p{1cm}>{\centering}p{1cm}>{\centering}p{1cm}|}
\hline 
$Par.$  & $h$  & $\Omega_{m,0}$  & $\Omega_{b,0}$  & $n$  & $\Omega_{DE}$  & $w_{0}$  & $w_{1}$  & $\gamma_{0}$  & $\gamma_{1}$  & $\sigma_{8,0}$\tabularnewline
\hline 
$1\sigma$  & 0.008  & 0.004  & 0.0008  & 0.03  & 0.009  & 0.07  & 0.31  & 0.19  & 0.49  & 0.03\tabularnewline
\hline 
$1\sigma$  & 0.008  & 0.004  & 0.0008  & 0.02  & 0.009  & 0.07  & 0.28  & 0.03  & -  & 0.007\tabularnewline
\hline 
\end{tabular}

\caption{\label{tab:Errors-on-cosmological}Uncertainties on cosmological parameters, collectively indicated 
as {\it Par.}
In the second row $1\sigma$ values refer to 68\% probability regions computed marginalizing
over all other parameters while in the third row values are obtained
through marginalization after fixing $\gamma_{1}$. }
\end{minipage}
\end{table*}



\begin{table*}
\centering %
\begin{minipage}[c]{170mm}%
\begin{tabular}{|>{\centering}p{0.7cm}|>{\centering}p{0.7cm}>{\centering}p{0.7cm}>{\centering}p{0.7cm}>{\centering}p{0.7cm}>{\centering}p{0.7cm}>{\centering}p{0.7cm}>{\centering}p{0.7cm}>{\centering}p{0.7cm}>{\centering}p{0.7cm}>{\centering}p{0.7cm}>{\centering}p{0.7cm}>{\centering}p{0.7cm}>{\centering}p{0.7cm}>{\centering}p{0.8cm}|}
\hline 
$z$  & $0.7$  & $0.8$$ $  & $0.9$  & $1.0$  & $1.1$  & $1.2$  & $1.3$  & $1.4$  & $1.5$  & $1.6$  & $1.7$  & $1.8$  & $1.9$  & $2.0$\tabularnewline
\hline 
$1\sigma$  & 0.023  & 0.022  & 0.022  & 0.023  & 0.025  & 0.027  & 0.029  & 0.031  & 0.034  & 0.038  & 0.039  & 0.041  & 0.044  & 0.046\tabularnewline
\hline 
$1\sigma$  & 0.010  & 0.011  & 0.011  & 0.012  & 0.013  & 0.014  & 0.016  & 0.016  & 0.018  & 0.02  & 0.021  & 0.022  & 0.024  & 0.026\tabularnewline
\hline 
\end{tabular}\caption{\label{bias_errors}In this Table we show the 1$\sigma$ uncertainties over
bias parameters in each redshift bin: in the second row $1\sigma$ values refer to uncertainties computed marginalizing over all other parameters
while in the third row values are obtained through marginalization
after fixing $\gamma_{1}$ and $\Omega_{K}$.}
\end{minipage}
\end{table*}

\section{Conclusions}\label{sec:conclusions}

The goal of this  Letter is to point out
explicitly that
a two-point statistics like the power spectrum
can provide independent constraints on
the cosmological parameters
 $\sigma_{8,0}$,  $\gamma$ 
and the galaxy bias, under the 
 quite general assumption that the 
 growth rate of density fluctuation
 can be  parametrized as $ f=\Omega_{m}(z)^{\gamma}$
 and the measurement is performed at two (or more) different redshifts. 
 This result  is not surprising.
 In fact  it may appear even obvious, but to the
best of our knowledge it has  never been explicitly pointed out 
in the literature.
To assess how precisely these parameters can be determined 
after breaking the degeneracy we have performed a Fisher-Matrix
analysis and explored the case of the next generation Euclid redshift 
survey \citep{euclid_rb}. 
We found that these quantities could be
measured to the percent level. Constraints tighten considerably
if a flat universe is assumed or if $\gamma$ is taken to be time independent.
 We find that $\sigma_{8}$ and $\gamma$
can be measured to within $1\%,5\%$, respectively, while the bias
$b(z)$ can be measured to within $1-2\%$ (assuming flatness) in
every redshift bin.

This procedure can be applied to redshift surveys like 
 WiggleZ \citep{blake11} that are deep 
enough to provide independent redshift shells in which to measure
the power spectrum. Analyses of this dataset have provided
estimates of combinations $f\sigma_{8,0}G(z)$ and  $\beta(z)$
\citep{blake11}. We will investigate the possibility of using 
the WiggleZ to estimate  $\sigma_{8}$, $\gamma$ and $b(z)$
in a future paper.

\section{Acknowledgments}
We thank Will Percival for his comments on a preliminary version of this work and B. Garilli and the NISP simulations group for providing us with the 
predicted galaxy redshift distribution for Euclid, as presented in the Euclid 
Red Book, and the Euclid Consortium for allowing access to this 
information.
CDP acknowledges  the  support  provided by  INAF  through  a 
PRIN/2008/1.06.11.10 grant.
CDP also thanks the Institut f$\ddot{\textrm{u}}$r Theoretische Physik and the University of Heidelberg for the kind hospitality and Carmelita Carbone for useful discussions on this topic.
EB  acknowledges the support provided by
MIUR PRIN 2008 ``Dark energy and cosmology with large galaxy surveys'' and by
Agenzia Spaziale Italiana (ASI-Uni Bologna-Astronomy Dept. `Euclid-NIS' I/039/10/0). L.A. acknowledges
funding from DFG through the project TRR33 ``The Dark Universe''.



\end{document}